\documentclass[%
superscriptaddress,
twocolumn,
 amsmath,amssymb,
 aps,
 prp,
]{revtex4-2}

\usepackage{graphicx}% Include figure files
\usepackage{dcolumn}% Align table columns on decimal point
\usepackage{bm}% bold math
\usepackage{color}
\usepackage{hyperref}% add hypertext capabilities
\begin{document}

%\preprint{APS/123-QED}
\preprint{Preprint}

%EA
\title{Nonlinear electronic stopping of negatively-charged particles in liquid water}

%\title{Nonlinear effects in the electronic stopping of 
%\textcolor{red}{negative-charge}
%\textcolor{blue}{negatively-charged} \textcolor{red}{NK: its not only negative..} %particles in liquid water}
%\title{Nonlinear effects in the electronic stopping power of liquid water for negative and positive projectiles}
%electrons, protons, and muons}% Force line breaks with \\

\author{Natalia E. Koval}
 \altaffiliation[Corresponding author: ]{natalia.koval.lipina@gmail.com}
 \affiliation{CIC Nanogune BRTA, Tolosa Hiribidea 76, Donostia-San Sebasti\'an, Spain.}%Lines break automatically or can be forced with \\
%\affiliation{%


\author{Fabiana Da Pieve}
\altaffiliation[Currently working at the European Research Council Executive Agency (ERCEA). The information and views set out in this article are those of the author and do not necessarily reflect the official opinion of the ERCEA.]{}
\affiliation{Royal Belgian Institute for Space Aeronomy, Brussels, Belgium.
}%
%\affiliation{

\author{Bin Gu}
\affiliation{Atomistic Simulation Centre, Queen’s University Belfast, Belfast BT7 1NN, Northern Ireland, United Kingdom}%
\affiliation{Department of Physics, Nanjing University of Information Science and Technology, 210044, Nanjing, China}%

\author{Daniel Mu\~noz-Santiburcio}
\affiliation{Instituto de Fusi\'on Nuclear ``Guillermo Velarde'', Universidad Polit\'ecnica de Madrid, Spain}

\author{Jorge Kohanoff}
\affiliation{Atomistic Simulation Centre, Queen’s University Belfast, Belfast BT7 1NN, Northern Ireland, United Kingdom}
\affiliation{Instituto de Fusi\'on Nuclear ``Guillermo Velarde'', Universidad Polit\'ecnica de Madrid, Spain}%

\author{Emilio Artacho}
\affiliation{CIC Nanogune BRTA and DIPC, Tolosa Hiribidea 76, 20018 San Sebasti\'an, Spain.}
\affiliation{Theory of Condensed Matter, Cavendish Laboratory, University of Cambridge, 
             J. J. Thomson Avenue, Cambridge CB3 0HE, United Kingdom.}
%\affiliation{Donostia International Physics Center DIPC, P. Manuel de Lardizabal 4, 
%             San Sebasti\'an, Spain.}
\affiliation{Ikerbasque, Basque Foundation for Science, 48011 Bilbao, Spain.}


\date{\today}% It is always \today, today,

\begin{abstract} 
  We present real-time time-dependent density-functional-theory 
calculations of the electronic stopping power for negative 
and positive projectiles (electrons, protons, 
antiprotons and muons) moving through liquid water.
After correction for finite mass effects,
the nonlinear stopping power obtained in this work is
significantly different from the previously
known results from semi-empirical calculations 
based on the dielectric response formalism. 
  Linear--nonlinear discrepancies are found both in
the maximum value of the stopping power and the Bragg
peak's position.
  Our results indicate the importance of the nonlinear 
description of electronic processes, particularly for 
electron projectiles, which are modeled here as classical
point charges. 
  Our findings also confirm the expectation 
that the quantum nature of the electron projectile should
substantially influence the stopping power around the 
Bragg peak and below.
\end{abstract}

\maketitle

%\tableofcontents

\section{\label{sec:intro}Introduction}

  The problem of electronic stopping of charged particles 
in matter is of continuing interest in fundamental science 
and in many applied research areas.
  In particular, an accurate description of the damage caused 
by energetic protons and electrons in biological tissue is 
crucial for hadron radiotherapy of cancer ~\cite{jackel,tommasino} 
and space exploration~\cite{ewa,dur_cuc,foray,walsh}.
  The effect of ionizing radiation on the DNA components, 
the main subject of radiobiology, is an active field of 
research in which the electronic effects are yet to be 
understood~\cite{papas}.

An energetic particle moving through biological
matter continually transfers energy to the target nuclei and electrons. 
The rate at which the projectile loses energy to the target per unit length of trajectory is called the stopping 
power, usually separated in electronic and nuclear contributions. Nuclear stopping power is primarily important for heavy projectiles with relatively low kinetic energies. Conversely, for fast projectiles the most important energy loss mechanism is electronic stopping. In this work, we study the impact of fast light projectiles and thus only focus on the electronic stopping power (ESP), which constitutes the first stage of the radiation damage process.

For decades, researchers have been using semi-empirical methods based
on the dielectric response formalism to study radiobiological effects of 
ionizing radiation~\cite{papas}, in particular, to calculate the ESP~\cite{FRANCIS20112307,emfi2005bis,Emf2009}.
Liquid water is commonly used as a target since semi-empirical methods rely
on experimental data not available for DNA components.
At high proton and electron velocities, the ESP can be well described by 
linear-response theory. However, the linear description is no longer 
applicable when the particles travel at intermediate and low velocities
(around the Bragg peak and lower). Moreover, for light particles at sufficiently low velocities, {\it e.g.}, electrons towards the end of the track, nuclear stopping and quantum effects become relevant.

Recent developments in density functional theory (DFT) and its
time-dependent extension (TDDFT) have advanced significantly the description 
of the electronic stopping processes in materials in the whole range of
velocities~\cite{Correa2018}. Most of the studies are focused on 
solid-state materials~\cite{PhysRevLett.99.235501,PhysRevLett.108.225504,
Schleife2015,PhysRevB.91.125203,Maliyov2018,Koval:rsos.200925},
although, some \emph{ab initio} simulations for protons in liquid water 
became available in recent years. Real-time (RT-) TDDFT calculations of the
proton stopping in water, ice, and water vapor provide accurate results and show a 
quantitative agreement with available experiments~\cite{PhysRevB.94.041108,Gu2020,Gu2022}. 

For electrons in water, no studies addressing the nonlinearity of the electronic stopping
processes are available to date. However, understanding the nonlinear
effects in the interaction of electrons with water is of great 
importance for benchmarking semi-empirical methods and for providing 
access to the low-energy region in which the dielectric response formalism 
is not expected to be valid~\cite{Shukri}.
Hence, in this work, we present a detailed analysis of the nonlinear effects in the ESP for negative and positive projectiles representing electrons, protons, and muons in water calculated using RT-TDDFT.
We compare our results with dielectric-response calculations and other available data such as SRIM~\cite{srim} and ESTAR~\cite{estar}.  
We analyse as well the effect of the projectile charge, the so-called
Barkas effect~\cite{barkas1963resolution}, on the electronic stopping.

\section{\label{sec:method} Methodology and numerical details}

%\subsection{\label{sec:rt-TDDFT} Real-time time-dependent density functional theory}

We used the RT-TDDFT implementation of the open-source {\sc Siesta} code
~\cite{PhysRevB.66.235416,PhysRevB.91.125203} to evolve the electronic 
orbitals in time,
as implemented in version master-post-4.1-264,  
available at https://gitlab.com/npapior/siesta/-/tree/geometry-motion.
  In SIESTA, the time-dependent Kohn-Sham (KS) equations are 
solved by real-time propagation of the KS orbitals using the 
Crank-Nicholson scheme~\cite{crank_nicolson_1947} as recently implemented by Halliday and Artacho ~\cite{JESS2019,PhysRevResearch.3.043134}. The new implementation 
replaces the Sankey integrator \cite{sankey2001} 
known to be problematic at high energies \cite{artacho2017}.
The forces on the nuclei of the target atoms and on the projectile itself 
are disregarded in the time propagation, thereby describing electron
dynamics with frozen host nuclei and a constant velocity projectile, as 
done in many similar studies~\cite{Gu2020,PhysRevLett.99.235501,PhysRevLett.108.225504,PhysRevB.91.125203}.
In this way we can separate the electronic and nuclear 
contributions to the total stopping and only consider the ESP with a clear
velocity dependence.
The ESP, $S_e=dE_{\mathrm{KS}}(x)/dx$, is obtained from a linear fit of the KS total electronic energy $E_{\mathrm{KS}}(x)$ with respect to the projectile displacement $x$, along the constant-velocity path.
 This expression is known to give the 
correct value of $S_e$ within the density-functional theory
defined by the chosen exchange-correlation functional, as
long as it is an adiabatic one~\cite{Correa2018, 
Halliday2022}

The water samples and the projectile trajectories are as of
Bin Gu et al.~\cite{Gu2020}. The simulation cell consisted of 203 
water molecules. A total of seven trajectories
%along the $x-$axis 
were considered for each projectile. Bin Gu et al.~\cite{Gu2020} 
showed that with a limited number of rigorously chosen trajectories it is 
possible to reproduce accurately the statistically averaged experimental ESP.

The electronic ground state of the target
water sample %(Fig.~\ref{Fig:water})
was calculated
using the static DFT implementation of the {\sc Siesta}
code~\cite{Soler2002} using periodic boundary conditions. For each trajectory, 
the projectile was placed at the initial position in DFT calculations.
%The ground state eigenstates of the liquid water sample were obtained by solving the KS equations self-consistently.
We used the generalized gradient approximation (GGA) in the Perdew, Burke,
and Ernzerhof (PBE) form for the exchange-correlation 
functional~\cite{PhysRevLett.77.3865}. Norm-conserving Troullier-Martins
~\cite{PhysRevB.43.1993} 
relativistic pseudopotentials were used to represent the core electrons.
The valence electrons were represented by a triple-$\zeta$ polarized (TZP)
basis set of numerical atomic orbitals with the default energy shift of 
0.02 Ry~\cite{Artacho1999}. The electronic Brillouin zone was
sampled at the $\Gamma$-point. The real-space grid was determined by a
plane-wave cutoff of 1000 Ry.
The KS states were then evolved in time by performing the RT-TDDFT calculations for each projectile moving with different velocities using the time step of 1 attosecond.
The convergence of $S_e$ with respect to 
the time step and the Brillouin zone sampling was tested in Ref.~\cite{JESS2019}.

\begin{figure}[t]
\includegraphics[width=0.48\textwidth]
{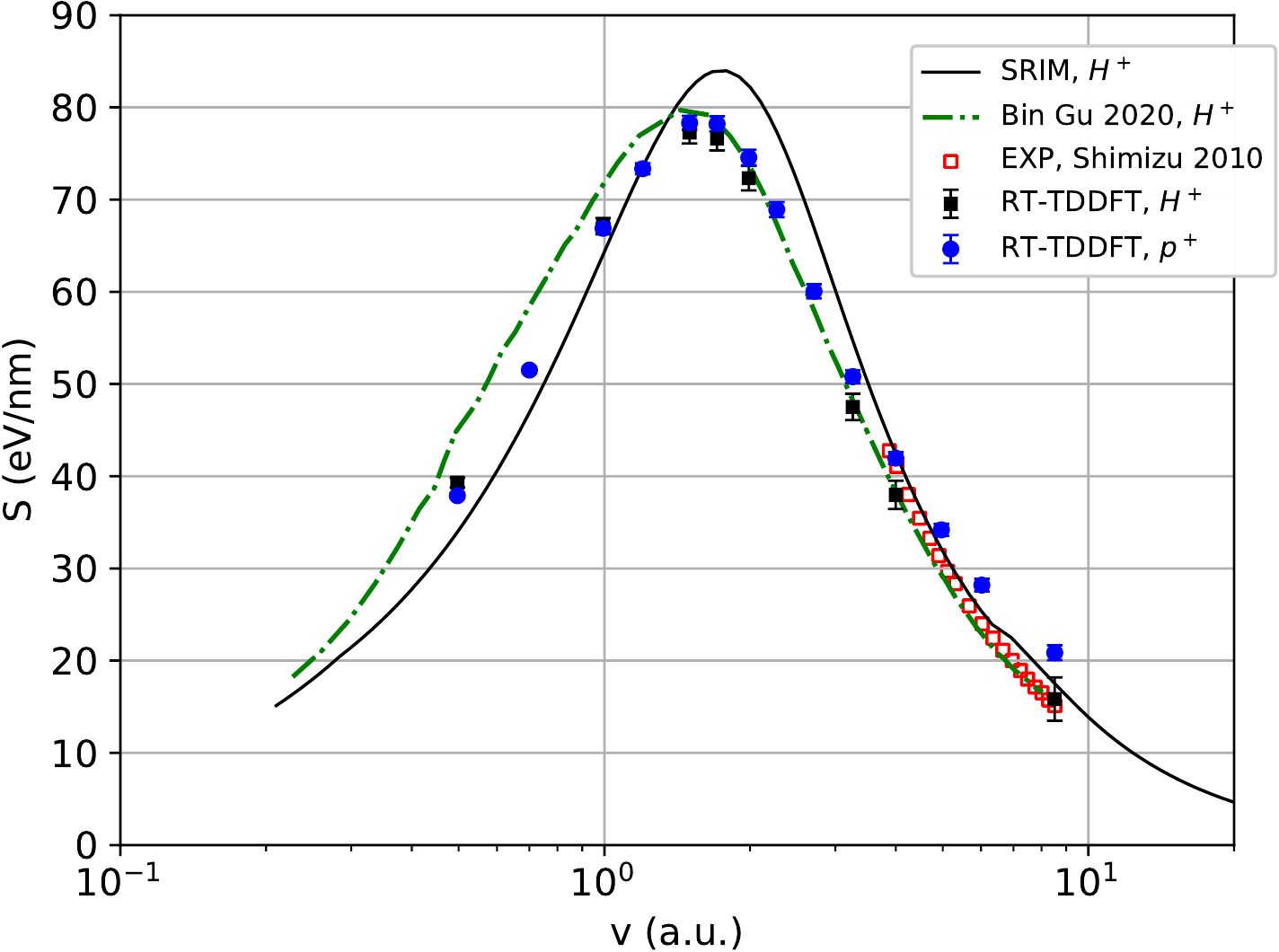}
\caption{\label{Fig:Stop_pr} ESP for a proton in liquid water 
as a function of velocity. Our results are presented for both the 
point charge $p^+$ and the hydrogen ion H$^+$ (full symbols) and 
are compared with the SRIM data~\cite{{srim}} (solid line), 
the RT-TDDFT results from Bin Gu et al.~\cite{Gu2020} for H$^+$ (dash-dotted line), and 
experimental data from Shimizu et al.~\cite{SHIMIZU20101002} (empty symbols). The error bars of the 
RT-TDDFT results depict the accuracy of the linear fit.}
\end{figure}

The point-charge projectiles were modeled via a spherical Gaussian charge distribution, using 
a {\sc Siesta} feature that allows the modeling of charged 
objects of different shapes.~\cite{C5CP04613K}.
The parameters
defining the Gaussian charge distribution 
were determined from the comparison of the ESP 
for a proton projectile moving with the velocity of 1.71 a.u. and modeled both as an explicit hydrogen atom 
and via a Gaussian charge distribution \footnote{The additional electron 
implied by a H atom vs a proton does not perceptibly affect the 
final result in a large enough sample.},
taking the latter as a reference. We determined that the Gaussian positive 
charge ($p^+$) distribution given by a width 
of $\sigma=$ 0.05 {\AA} and 
a cutoff of 0.5 {\AA} leads to the same stopping power 
within 2 \%
as the hydrogen projectile (H$^+$), as can be seen in 
Fig.~\ref{Fig:Stop_pr}. The basis set for the projectile was 
provided by a ghost hydrogen atom. We used a triple-$\zeta$ 
doubly-polarized (TZ2P) basis set on the projectile with cut-
off radii $r(\zeta_1) = 8.80$, $r(\zeta_2) = 6.853$, and 
$r(\zeta_3) = 0.50$ Bohr 
for electrons to adapt to
the narrow Gaussian distribution. 
The dependence of $S_e$ on the Gaussian charge 
width is discussed in the Appendix.

\begin{figure*}[t]
\includegraphics[width=0.99\textwidth]{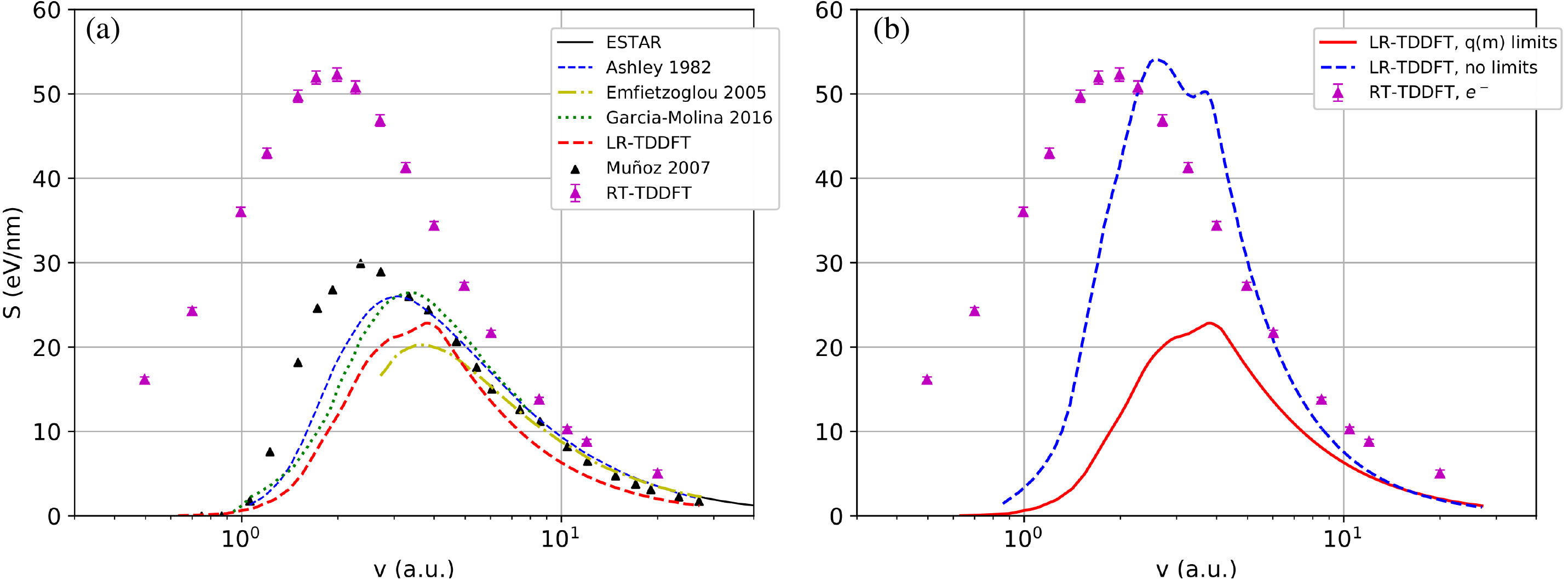}
%{Stopping-electron-average-SIESTAnew-TZP-ghost-basis-narrow-exp-crop.pdf}
%\includegraphics[width=0.48\textwidth]
%{Stopping-electron-average-siesta-vs-lr-tddft-no-limit-crop.pdf}
\caption{\label{Fig:Stop_el} (a) ESP for an electron in liquid water
as a function of velocity calculated with RT-TDDFT compared with the 
results obtained using the dielectric response formalism from Emfietzoglou et al.~\cite{emfi2005bis} (dash-dotted line), Gracia-Molina et al.~\cite{garcia2017_49} (dotted line) and with the LR-TDDFT stopping power~\cite{Koval_rsos2022} (dashed line). The semi-empirical data points from Mu\~noz et al~\cite{PhysRevA.76.052707} are obtained by converting the mass stopping power for gas phase to ESP by assuming the density of 1 g/cm$^3$. ESTAR 
data~\cite{estar} (thin solid line, on the most right) and Ashley et al.~\cite{Ashley1982} also based on dielectric response (thin dashed line) are also presented for comparison. (b) Comparison of ESP obtained from RT-TDDFT (full symbols) and from LR-TDDFT using the mass-dependent integration limits for $q$ (see~\cite{Koval_rsos2022}) (solid line) and without such limits (dashed line). See text for explanation.}
\end{figure*}

The agreement between our results and Bin Gu et al.~\cite{Gu2020} for H$^{+}$ is very reasonable although, at low proton velocities, our ESP is slightly lower than the reference result. The discrepancy may be
associated with the differences in the pseudopotentials 
and basis sets used in our {\sc Siesta} calculations, versus 
the ones used with the {\sc cp2k} code
by Bin Gu et al.~\cite{Gu2020}, and with the fact that they performed 
all-electron calculations.
The ESP for $p^+$ and H$^+$ from {\sc Siesta} RT-TDDFT obtained 
in this work and shown in Fig.~\ref{Fig:Stop_pr}, are in close 
agreement with each other except for the highest velocities, where 
the $p^+$ results are slightly higher than the rest of the data sets.
The discrepancy mainly comes from the use of different basis sets for the two projectiles, the TZP basis set for the H$^+$ and the TZ2P one for the ghost atom moving with $p^+$. The stopping power is more sensitive to the choice of the basis set at high velocities for both projectiles as our test calculations have shown.
As stated before, the Gaussian charge demands more basis than the pseudised proton.

\section{\label{sec:res} results and discussion}

Figure \ref{Fig:Stop_el}(a) shows the comparison of our 
RT-TDDFT stopping 
power for a negative point charge (an electron) with the ESP obtained using 
semi-empirical methods by Garcia Molina et 
al.~\cite{garcia2017_49}, Emfietzoglou et 
al.~\cite{emfi2005bis} and Mu\~noz et al.~\cite{PhysRevA.76.052707} (theory combined with experiment), as well as to a dielectric model developed by Ashley et al.~\cite{Ashley1982} and ESTAR~\cite{estar}. 
The LR-TDDFT result also presented in Fig.~\ref{Fig:Stop_el}(a) is obtained from the \emph{ab initio} energy loss function~\cite{Koval_rsos2022}.
The position of the Bragg peak is significantly
different. The linear results show a maximum at around $3-4$ a.u.
(energy of the order of $100-200$ eV),
while the RT-TDDFT gives us the maximum stopping at the velocity
of 2 a.u. ($\sim 50$ eV). Slightly closer to ours is the position of the Bragg peak obtained by Mu\~noz et al.~\cite{PhysRevA.76.052707}. These data points are calculated using experimental cross-sections for gas-phase water.

Apart from the large discrepancy between the position of the Bragg peak in the 
linear and nonlinear stopping power for electrons 
observed in Fig.~\ref{Fig:Stop_el}(a), the maximum 
value of the ESP is also drastically different.
The linear results largely underestimate the stopping power in
a wide range of velocities as compared to our \emph{ab-initio} nonlinear ESP.
A significant part of the discrepancy, however, 
stems from finite-projectile-mass effects, as follows.

Since we use a constant velocity approximation for the 
electron in our RT-TDDFT calculations, this implies that its mass is 
infinite. In the linear calculations based on the integration of the 
electron energy loss function, although the approximation is built for a constant velocity perturbation, the electron mass is accounted for in the integration limits for the momentum transfer $q$ coming from the energy and momentum 
conservation~\cite{emfi2005bis}. Removing such integration limits for $q$, and thus integrating from zero to infinity over
the momentum transfer, we obtain a much higher ESP, as can be seen in 
Fig.~\ref{Fig:Stop_el}(b). Both RT-TDDFT and LR-TDDFT stopping power for infinite electron mass have peaks of
similar height. However, the nonlinear effect is still noticeable
as the Bragg peak position of the RT-TDDFT ESP is shifted by approximately 1 a.u. of velocity. The comparison on Fig.~\ref{Fig:Stop_el}(b)
emphasises that a constant velocity is a crude approximation for
an electron. Further studies are required to account for 
the electron mass in the RT-TDDFT calculations.

\begin{figure*}[t]
\includegraphics[width=0.98\textwidth]
{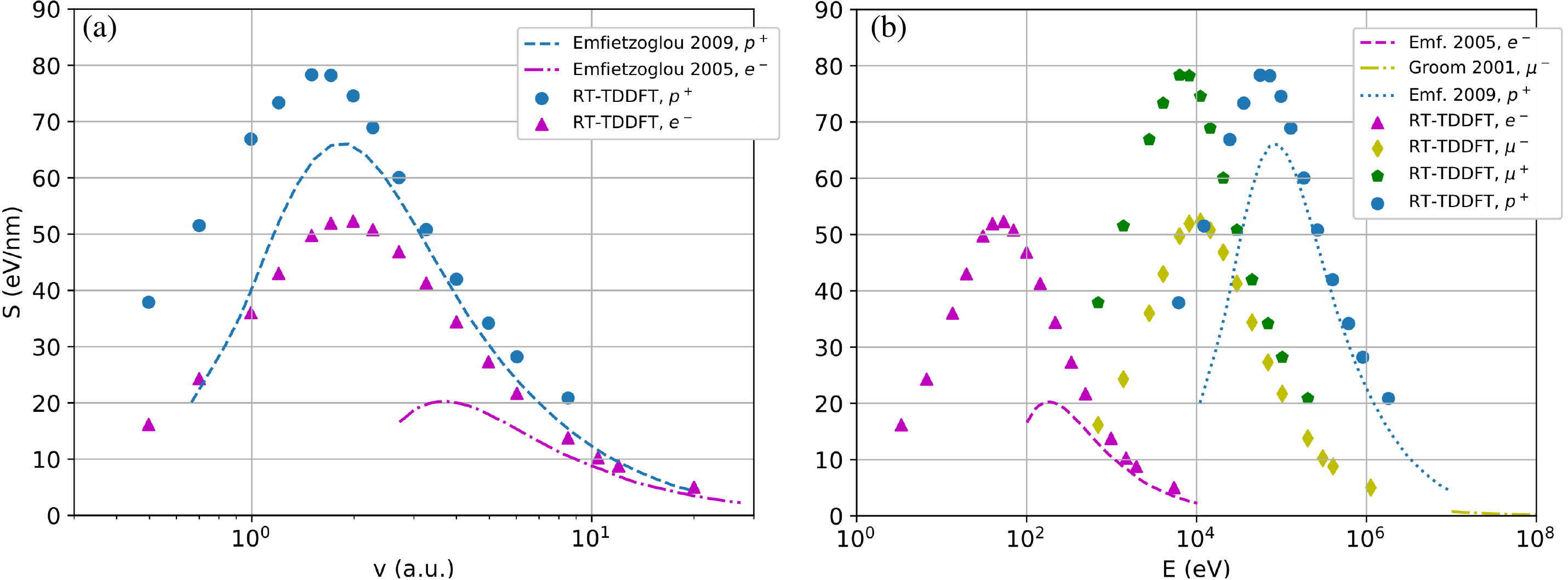}
%{Stopping-e-p-average-vs-Emfietzoglou-crop.pdf}
%\includegraphics[width=0.48\textwidth]
%{Stopping-e-p-mu-average-vs-Emfietzoglou-energy-crop.pdf}
\caption{\label{Fig:e-p-ener} (a) ESP for a proton and an electron in liquid water as a function of velocity obtained in this work within RT-TDDFT (full symbols) compared with the dielectric response results from Refs.~\cite{Emf2009,emfi2005bis} (lines). (b) ESP for an electron, a muon (both negative and positive), and a proton in liquid water as a function of the projectile kinetic energy from RT-TDDFT (full symbols) compared with the dielectric response results for $e^-$~\cite{emfi2005bis} (dashed line) and $p^+$~\cite{Emf2009} (dotted line), and Bethe-Bloch formula for $\mu^-$~\cite{GROOM2001183} (dash-dotted line).}
\end{figure*}

The maximum of the ESP for $p^+$ is higher than for $e^-$ obtained with RT-TDDFT, a phenomenon known as the Barkas effect~\cite{Barkas_1963} (Fig.~\ref{Fig:e-p-ener}(a)). Notice that in these calculations, the mass of the particle is not taken into account. Hence, the same ESP corresponds to a proton and a positron. The position of the maxima (the Bragg peaks), however, is very similar ($v=1.7$ a.u. and 2 a.u. for 
proton and electron, respectively). This is not true in the case of the linear results of Emfietzoglou et al., in which the Bragg peak is observed at 2 a.u. for the proton and at 4 a.u. for the electron projectile. Overall, the linear--nonlinear discrepancy is much more pronounced for electron projectiles.

Figure \ref{Fig:e-p-ener}(b) shows the ESP for an electron, a negative and a positive muon, and a proton as a function of the projectile kinetic energy. For the positive (negative) muon, 
we used the results of the proton (electron) scaling the kinetic energy taking into
account the muon mass $M_\mu=206.768~m_e$~\cite{nist}. The Bragg peak is at energies of  
$\sim50$ eV, $10$
keV, and $90$ keV for the electron, muons, and proton, respectively.
The Bragg peak energies scale linearly with the masses of the three particles (e.g. for a proton vs muon $M_p/M_\mu=8.88m_e$)
as expected, since their only mass dependence
arises from the velocity-energy conversion, but which is not 
the case for the linear results of Ref.~\cite{Emf2009,emfi2005bis}.
For a negative muon, the Bethe-Bloch result~\cite{GROOM2001183} is only available at energies above $10^7$ eV, out of range of our calculations.

\section{Conclusions}

In conclusion, we presented the electronic stopping power 
for negative and positive projectiles in liquid water obtained 
with RT-TDDFT and compared to linear results available in the 
literature. 
Correcting for projectile mass effects,
the nonlinear effects have been shown to be
prominent in the electron-water interaction given the large difference between the linear and nonlinear ESP. This effect, however, has to be verified by calculations considering the quantum nature of the external electron and accounting properly for its finite mass. 

\section{Acknowledgements}

%We thank Nick Papior for extending the implementation of the {\sc siesta} code.
We acknowledge funding from the Research Executive Agency under 
the European Union's Horizon 2020 Research and Innovation program (project
ESC2RAD: Enabling Smart Computations to study space RADiation effects,  Grant  Agreement 776410). JK was supported by the Beatriz 
Galindo Program (BEAGAL18/00130) from the Ministerio de Educaci\'on y 
Formaci\'on Profesional of Spain, and by the Comunidad de Madrid through 
the Convenio Plurianual with Universidad Polit\'ecnica de Madrid in its 
line of action Apoyo a la realizaci\'on de proyectos de I+D para 
investigadores Beatriz Galindo, within the framework of V PRICIT (V Plan 
Regional de Investigaci\'on Cient\'ifica e Innovaci\'on Tecnol\'ogica). EA
acknowledges the funding from Spanish MINECO through
grant FIS2015-64886-C5-1-P, and from Spanish MICIN
through grant PID2019-107338RB- C61/AEI/10.13039/501100011033, 
as well as a Mar\'{\i}a de Maeztu award to Nanogune, Grant 
CEX2020-001038-M funded by MCIN/AEI/ 10.13039/501100011033.
We are grateful for computational resources provided by the Donostia International Physics Center (DIPC) Computer Center and Barcelona Supercomputer Center (HPC grants FI-2021-2-0037 and FI-2021-3-0030).

\appendix*

\begin{figure}
\includegraphics[width=0.48\textwidth]
{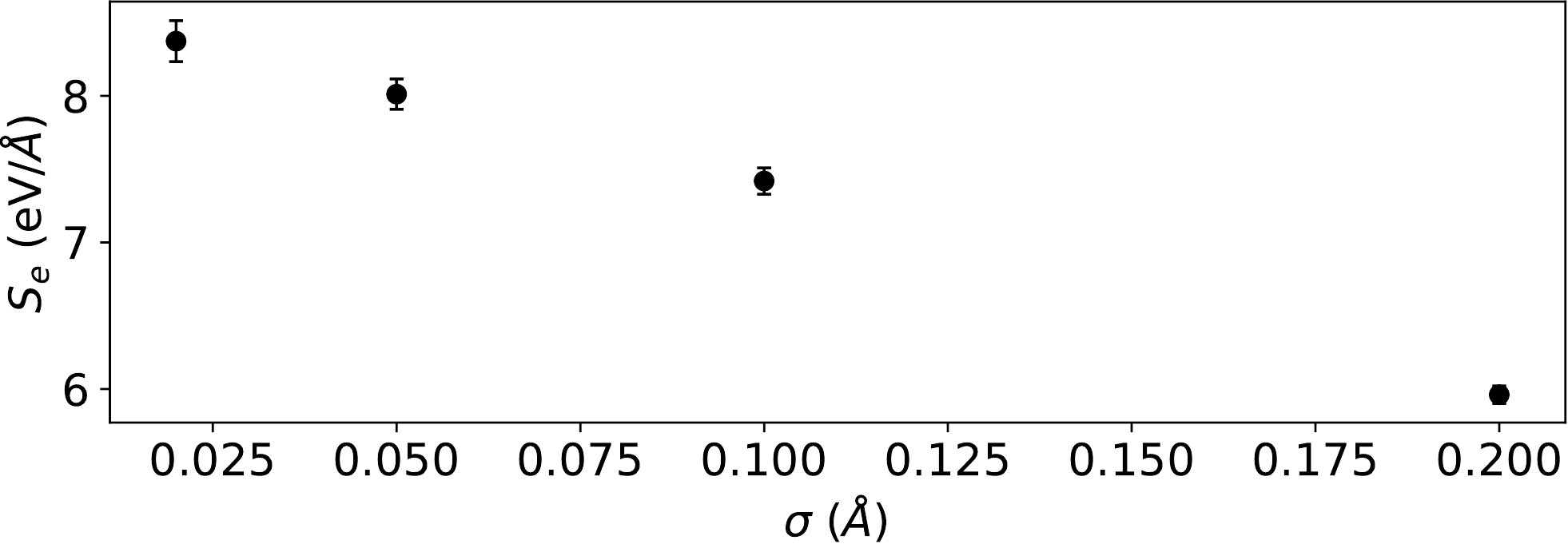}
\caption{\label{Fig:test-gaus} Electronic stopping power (in eV/\AA) 
for a proton in liquid water as a function of the width $\sigma$ 
(in \AA) of the spherical Gaussian 
% standard deviation $\sigma$ 
used to represent the proton charge, for a 
proton velocity of 1.71 a.u. The error bars of the 
RT-TDDFT results depict the accuracy of the linear fit
for extracting the stopping power.
%\textcolor{blue}{\sc EA: If easy I would remove the key box from the figure.}
}
\end{figure}

\section{Projectile charge width}

As already known from earlier work~\cite{PhysRevLett.99.235501},
the electronic stopping power depends on the smoothening of the 
Coulomb interaction of the projectile with the system electrons
at short distances.
Such smoothening is performed both when using pseudopotentials
and when substituting a point charge by a charge distribution.
 Our test results have shown that indeed the width of the Gaussian 
affects the stopping power in the calculations in this work. 
  Namely, as could be expected,
the ESP increases as the 
Gaussian becomes narrower (see Fig.~\ref{Fig:test-gaus}).

The limit of zero width does not 
represent a convergence target, 
however, since the projectile is not a classical point
charge. It could be argued that such a width should scale
with the de Broglie wave length.

From a technical point of view, such a width should not be 
smaller~\cite{PhysRevLett.99.235501} than the discretization of 
real space used to compute the Hamiltonian matrix elements, and 
specified by the plane-wave energy cutoff~\cite{Soler2002}, of
1000 Ry in this work, which implies a half wavelength of 0.1 Bohr.
We chose to use the same Gaussian width for the whole range of 
velocities, obtained from reproducing the Bragg peak values obtained
with the explicit all-electron calculations in Ref.~\cite{Gu2020}
for the electronic stopping power for protons.
We then used the same width to model the negatively charged 
Gaussian-distributed projectiles.

%\clearpage
\bibliography{biblio}% Produces the bibliography via BibTeX.

\end{document}